\documentclass[natbib,smallextended]{svjour3}                     
\smartqed  
\usepackage{graphicx}
\usepackage{amsmath}
\usepackage{amsfonts}
\usepackage{amssymb}
\usepackage{enumerate}
\usepackage{verbatim}
\usepackage{subcaption}
\usepackage{color}
\newcounter{algno}
\newenvironment{algorithm}[1]{
\bigskip
\noindent
\rule{1\textwidth}{1mm}

\begin{footnotesize} 
{\tt #1:}
}
{
\end{footnotesize}
\noindent
\rule{1\textwidth}{1mm}
}

\def\R{{\mathbb R}}

\def\Z{{\mathbb Z}}

\usepackage[xcolor]{changebar}
\cbcolor{blue}
\def\bcb{}
\def\ecb{}
\usepackage{color}

\begin{document}

\title{Dynamic detection of anomalous regions within distributed acoustic sensing data streams using locally stationary wavelet time series \thanks{This work has been partially supported by the Engineering and Physical Sciences Research Council under grants: EP/L015692/1 and Shell Research Ltd.}
}
\subtitle{}

\titlerunning{Dynamic detection of anomalous regions within DAS data streams}        

\author{Rebecca E. Wilson         \and
        Idris A. Eckley \and
        Matthew A. Nunes \and
        Timothy Park 
}

\authorrunning{R. Wilson et al.} 

\institute{
 Rebecca E. Wilson \at
  STOR-i Centre for Doctoral Training, Lancaster University, LA1 4YF, United Kingdom
           \and
           Idris A. Eckley \at Department of Mathematics and Statistics, Lancaster University, LA1 4YF, United Kingdom \\ \email{i.eckley@lancaster.ac.uk}
           \and
           Matthew A. Nunes \at School of Mathematics, University of Bath, BA2 7AY, United Kingdom 
           \and
           Timothy Park \at Statistics and Data Science, Shell Global Solutions International BV, Amsterdam, The Netherlands
}

\date{Received: date / Accepted: date}

\maketitle

\begin{abstract}
Distributed acoustic sensing technology is increasingly being used to support production and well management within the oil and gas sector, for example to improve flow monitoring and production profiling. This sensing technology is capable of recording substantial data volumes at multiple depths within an oil well, giving unprecedented insights into production behaviour. However the technology is also prone to recording periods of anomalous behaviour, where the same physical features are concurrently observed at multiple depths. Such features are called `stripes' and are undesirable, detrimentally affecting well performance modelling. This paper focuses on the important challenge of developing a principled approach to identifying such anomalous periods within distributed acoustic signals. We extend recent work on classifying locally stationary wavelet time series to an online setting and, in so doing, introduce a computationally-efficient online procedure capable of accurately identifying anomalous regions 
within 
multivariate time series. 
\end{abstract}

\section{Introduction}\label{intro}

The ability to accurately analyse geoscience data at, or close to, real time is becoming increasingly important. For example, within the oil and gas sector this need can arise as a consequence of (i) the sheer volume of data now being collected and (ii) operational considerations.  It is this setting that we consider in this article, seeking to enable the rapid identification of certain anomalous features within Distributed Acoustic Sensing data obtained from an oil producing facility. Specifically, we seek to build on recent work within the non-stationary time series community to develop an approach that permits the online monitoring of these complex signals.  

The technology used to generate the data considered in this article, Distributed Acoustic Sensing (DAS), involves the use of a fibre-optic cable as a sensor in which the entire length of the fibre is used to measure acoustic or thermal disturbances. DAS originates from the defence industry where it is commonly used in security and border monitoring \citep{Owen}. Recently, the technology has been applied within the oil and gas industry, for example in pipeline monitoring and management \citep{Williams,Mateeva}. 
The use of DAS to monitor production volumes and composition within a well requires the installation of a fibre-optic cable along the length of the well combined with an interrogator unit on the surface \citep{paleja2015velocity}. This unit sends light pulses down the cable and processes the back-scattered light. The installation of such technology has become popular as it can be a cost effective way to obtain continuous, real-time and high-resolution information.

When monitoring the behaviour of wells it is important to be able to detect unusual occurrences, including potential corruptions of the data. 
Striping is one particular form of corruption that can have a particularly deleterious effect, rendering data potentially unusable in a specific time region. Stripes are characterised by sudden, and distinctive changes in the structure of the signal over time, see \cite{Mateeva} and \cite{ellmauthaler2017noise} for examples. These features can be present simultaneously across all channels or only apparent across a subset of channels, for example from the surface to a set depth within the well. Crucially, the occurrence of stripes simultaneously at different locations indicates that these features are not physical. Instead stripes can occur for a number of reasons, including a disturbance of the fibre-optic cable near the unit, or problems with the electronics due to the high sampling rate.

\begin{figure}[!htbp] 
\centering 
\begin{subfigure}[h]{0.8\textwidth}
 \includegraphics[width=104mm]{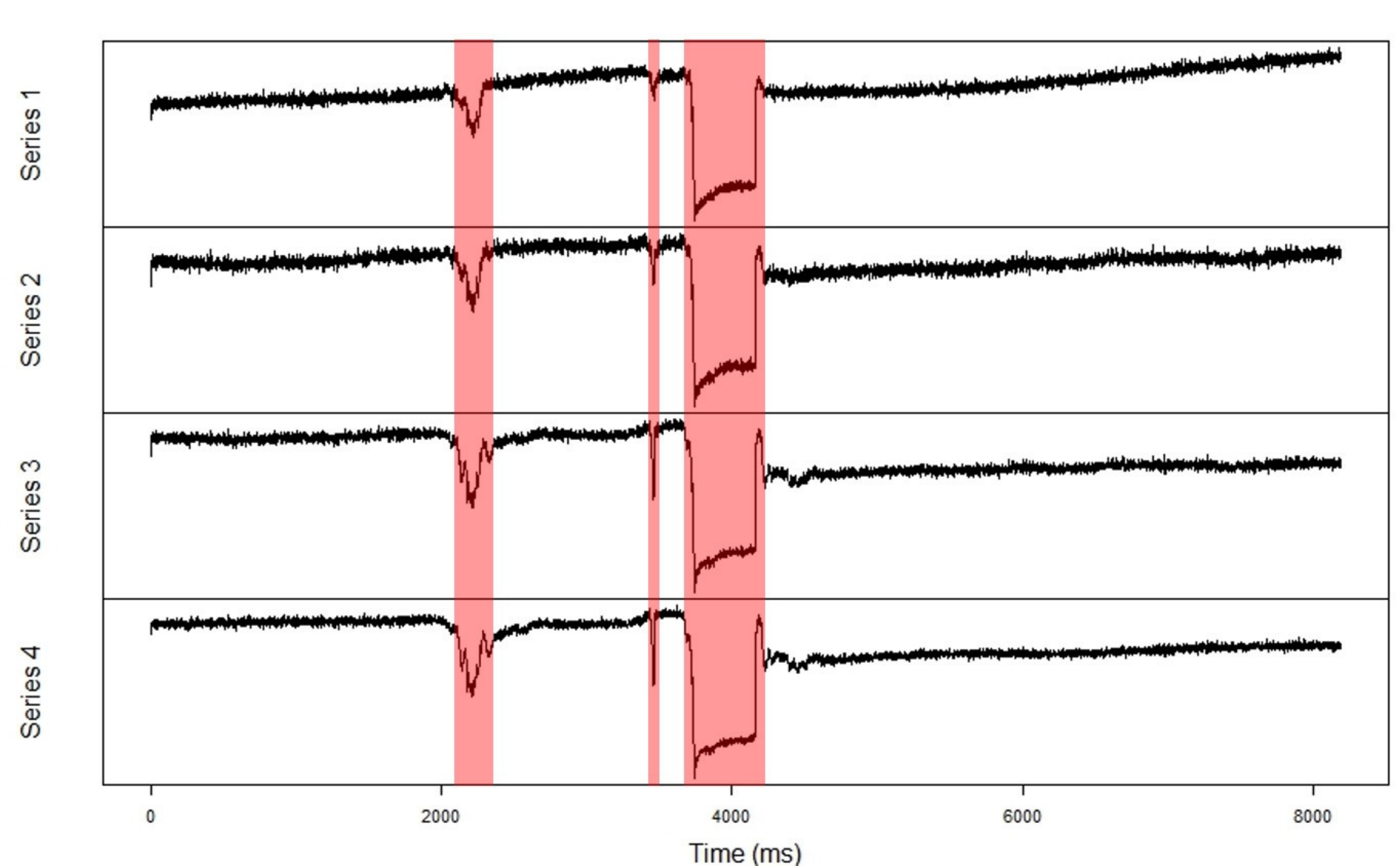} \label{fig:fig1a} \caption{}\end{subfigure} \bcb
 \begin{subfigure}[h]{0.8\textwidth}
 \includegraphics[width=107mm,height=70mm]{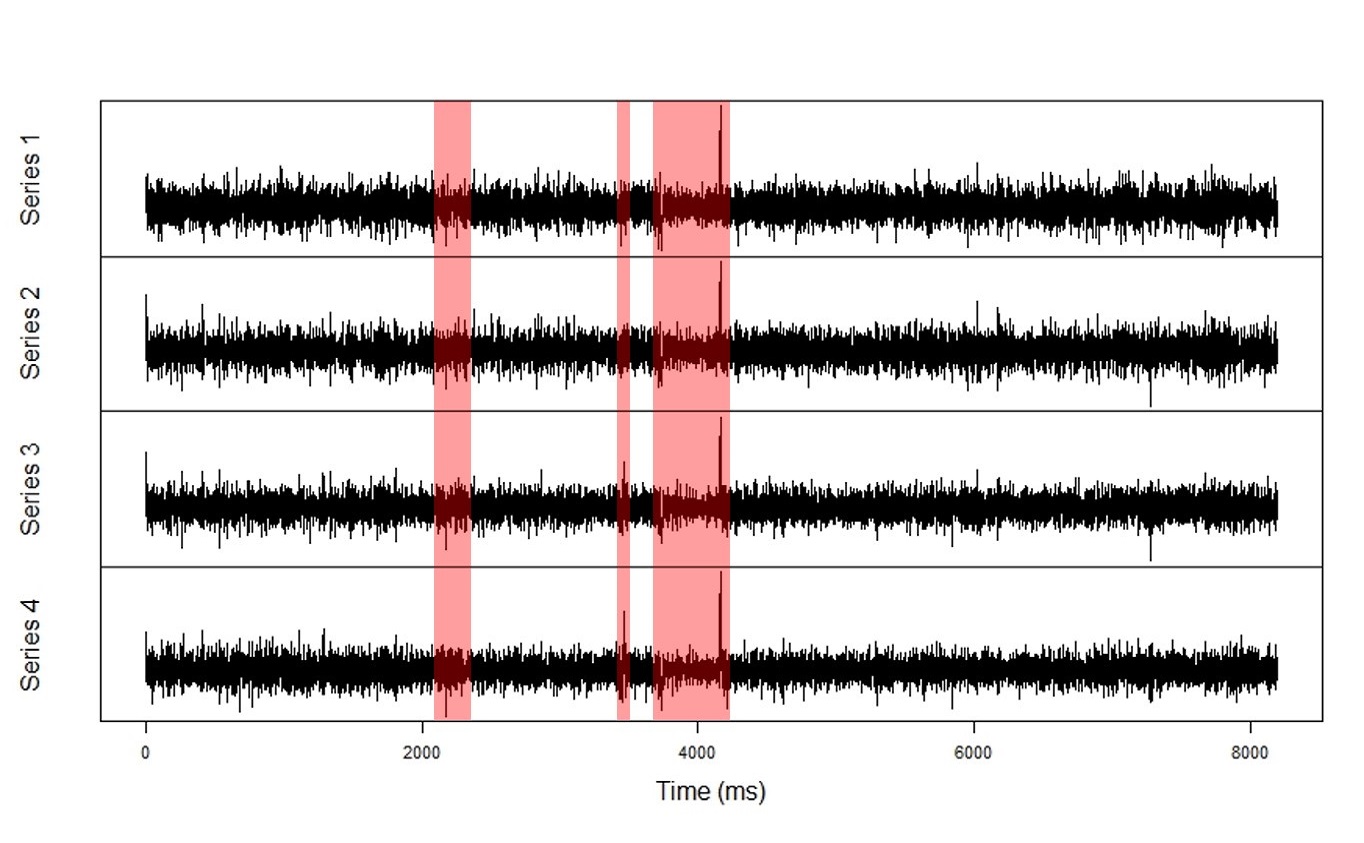} \caption{}\label{fig:fig1b} 
 \end{subfigure}\ecb
\caption{Time series plots of DAS amplitude at four different well depths over the same time period: (a) original series; (b) detrended series. The highlighted regions in (a) indicate three examples of striping.}
\label{fig:stripes}
\end{figure}

Visually, stripes can manifest themselves in a variety of ways. Some are visually obvious within the DAS data, such as the stripe that occurs at around $4000$ms in \bcb Figure \ref{fig:stripes}(a). Other occurrences can be more subtle, and therefore more challenging to detect. For example, the stripe could be a change in the second-order structure. \ecb Critically such features can make it difficult to carry out further analysis of the data, such as flow rate analysis. For this reason, there is significant interest in being able to detect regions of striping as soon as they occur, so that they can be removed whilst keeping as much of the original signal intact as possible. It is this challenge of dynamically detecting striping regions that motivates the work presented in this article.

\begin{figure}[!htbp] 
\centering 
\begin{subfigure}[h]{0.8\textwidth}
 \includegraphics[width=104mm]{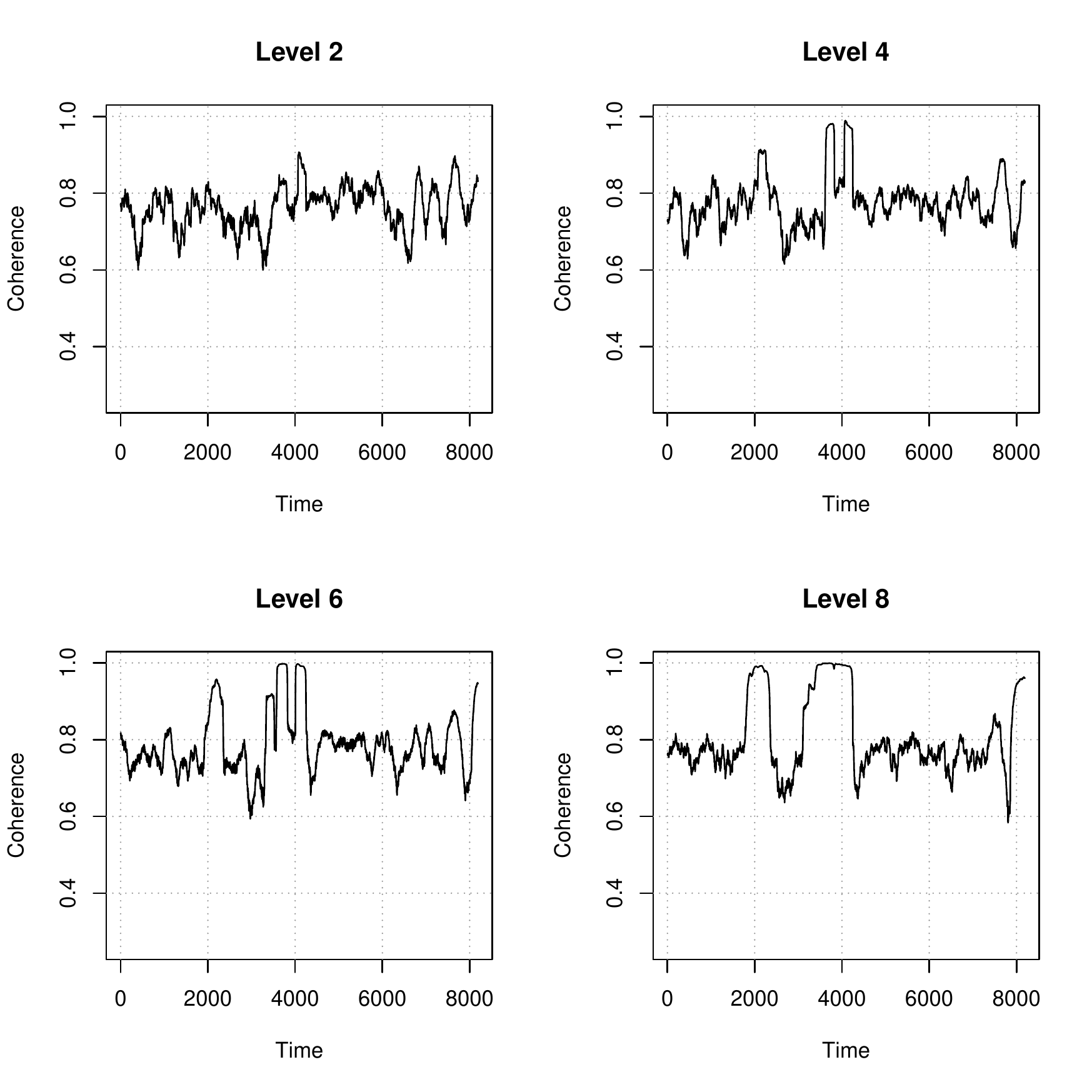} \caption{} \label{fig:fig2a} \end{subfigure} 
 \begin{subfigure}[h]{0.8\textwidth}
 \includegraphics[width=104mm]{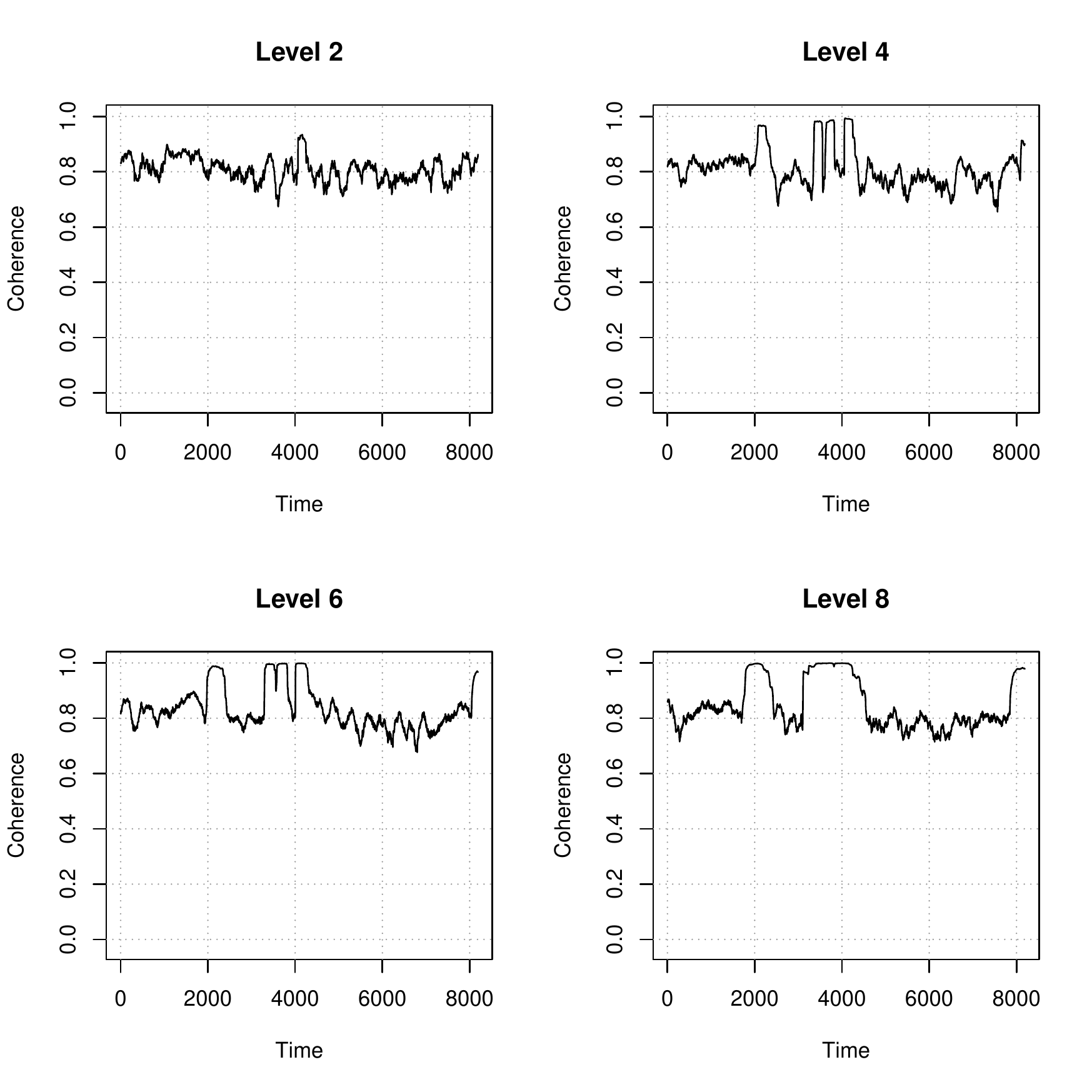} \caption{} \label{fig:fig2b} 
 \end{subfigure}
\caption{Hidden time-varying coherence structure of the DAS series in Figure \ref{fig:stripes} at selected wavelet scales (resolutions): (a) coherence between series 1 and 2; (b): coherence between series 3 and 4.}
\label{fig:stripecoh}
\end{figure}

There exist a variety of techniques for the classification of time series in the statistical and machine learning literature.  An exhaustive review is beyond the scope of this article, but popular classification methods include hidden Markov models (HMM) (see e.g. \cite{rabiner89:a,ephraim02:hidden, cappe09:inference}); support vector machines \citep{cortes95:support, muller01:an, kampouraki09:heartbeat}; Gaussian mixture models \citep{mclachlan04:finite, povinelli04:time, kersten14:simultaneous}; nearest neigbour classifiers \citep{zhang04:a,wei06:semi} and multiscale methods \citep{chan99:efficient, morchen03:time, aykroyd16:classification} to name but a few.  More recent contributions for large-scale (online) classification include the MOA machine learning framework \citep{bifet10:moa, read12:scalable}.  For a recent overview of classification in the time series context, see for example \cite{fu11:a}.  \bcb Dependent on the application being considered, one might adopt various modelling choices. For 
example, some classifiers have distinct advantages, such as simplicity of implementation, speed or suitability for 
massive online applications. \ecb  However many, such as GMM or SVM-based approaches, do not explicitly allow temporal dependence or are limited to a narrow class of series structure (HMMs), which is seen as crucial to classification of time series in the majority of realistic settings (see e.g. \cite{bifet13:pitfalls}).  Complex hidden dependence structure is typical of the DAS data studied in this article (see Figure \ref{fig:stripecoh}).

Our approach to the dynamic stripe identification problem builds on recent work within the time series literature. Wavelet approaches to modelling time series have become very popular in recent years, principally because of their ability to provide time-localised measures of the spectral content inherent within many contemporary data (e.g. \cite{KEJ2013, NAEK2015,  ChauvonSachs2015, NPES2017SeriesA}). 
This {\em locally stationary} modelling paradigm is flexible enough to represent a wide range of non-stationary behaviour 
and has also been extended to enable the modelling and estimation of multivariate non-stationary time series structures (e.g. \cite{SFJ2010} and \cite{Park}). Typically these settings assume that the data have already been collected, and are available for offline analyses.

The novel contribution in this article is to employ the MvLSW modelling framework of \cite{Park} to represent the DAS data, using a moving window approach, thereby extending previous work to the online dynamic classification setting.  This modelling framework allows us to classify multivariate time series with complex dependencies both {\em within} and {\em between} channels of the series, including those which exhibit visually subtle changes in behaviour over time.  Reusing data calculations allows us to also produce a computationally efficient nondecimated wavelet transform in the online setting.

Our article is organised as follows. Section \ref{sec:background} contains an overview of the Multivariate Locally Stationary Wavelet (MvLSW) model and existing dynamic classification method. In Section \ref{sec:method}, we describe the proposed online classification method. Section \ref{sec:examples} contains a simulation study evaluating the performance of the proposed classifier using synthetic data, further justifying the use of time-varying coherence as a feature for classification. A case study using an acoustic sensing dataset is then described in Section \ref{sec:case}, where we discuss the utility of the proposed classifier as a stripe detection method. Finally, Section \ref{sec:conclusions} includes some concluding remarks. 

\section{Wavelets and time series}\label{sec:background}

The problem of modelling and analysing non-stationary time series can be approached in a number of ways that often involve assuming the changing second-order structure adopts a time varying spectrum or autocovariance. Examples within the existing literature include the oscillatory processes \citep{Priestley}, Locally Stationary Fourier model \citep{LSF} and time-varying autoregressive processes \citep{dahlhaus1999nonlinear}. Due to the high frequency nature of acoustic sensing data, we focus our attention on wavelet-based methods such as the Locally Stationary Wavelet (LSW) processes, introduced by \cite{Nason00}. The use of wavelets allows for the time-scale decomposition of a signal using computationally efficient transform algorithms, whilst also allowing the structure to change over time.

Methods for the classification of non-stationary time series can broadly be divided into two categories; static or dynamic. Static classification approaches attempt to assign an entire test signal to a particular class. They differ in the way in which they choose to model the nonstationarity, including through Locally Stationary Fourier processes \citep{ST}, the smooth localised exponentials (SLEX) framework \citep{Huang} and wavelets \citep{Fry,Krz}. In contrast, dynamic classification approaches allow for the class assignment of the test signal to vary over time which allows for more flexibility in the classification and covers problems where the underlying nonstationarity is due to class switching. The method that we introduce is an online analogue of the dynamic classification approach of \cite{ParkDC} which looks to detect subtle changes in the dependence structure of a multivariate signal in a fast and efficient manner.

Before introducing our proposed method in Section \ref{sec:method},  we first outline some details of the locally stationary wavelet framework of \cite{Park} together with the (offline) dynamic classification method introduced in \cite{ParkDC} that forms the basis of our online approach.  We begin with some introductory concepts of wavelets. For a more comprehensive introduction to the area, see \cite{NasonTB} or \cite{Vidakovic}.

\subsection{Discrete wavelet transforms}\label{subsec:wavelets}

Succinctly, wavelets can be seen as oscillatory basis functions with which one can represent signals in a multiscale manner.  More specifically, for a function (signal) $f\in L^2(\R)$, we can write 
$$f(x) =\sum_{k\in\Z} c_{0,k} \phi_{0,k}(x) + \sum_{j\leq J}\sum_{k\in\Z} d_{j,k} \psi_{j,k}(x),$$ 
for scales $j$ and locations $k$, and where the wavelet $\psi_{j,k}(x) = 2^{-j/2}\psi(2^{-j}x-k)$ is a basis function formed as a dilation and translation of a ``mother'' wavelet $\psi$; scaling functions $\phi_{j,k}$ are similarly formed as dilated and translated versions of a father wavelet $\phi$.  The wavelet coefficients $d_{j,k}$ capture local oscillatory behaviour of the signal at a scale (frequency) $j$, whereas the scaling coefficients $c_{j,k}$ represent the signal's smooth (mean) behaviour at different scales.  More specifically, fine scale coefficients capture the local characteristics of a signal; coarse scale coefficients describe the overall behaviour of the signal.

\paragraph{The Discrete Wavelet Transform (DWT).} Computation of wavelet coefficients resulting from traditional wavelet transforms is performed using the so-called {\em Discrete Wavelet Transform} (DWT), first introduced by \cite{Mallat}.  The algorithm proceeds by alternately applying high- and low-pass filtering and decimation (subsampling) operations to the observed data.  

Let $\mathcal{H}:=\{h_{k}\}$ and $\mathcal{G}:=\{g_{k}\}$ be a low- and high-pass filter pair associated with a given wavelet, such as the quadrature mirror filters used in the construction of compactly supported wavelets introduced by \cite{daubechies}. Following \cite{NSDWT}, let $ \mathcal{D}_{0} $ denote the even decimation operator that selects every even-indexed element in a sequence, in other words $(\mathcal{D}_{0}x)_{l}=x_{2l}$. The detail coefficients of the DWT of a time series $X=\{x_{t}\}_{t=1}^{T}$ (where $T=2^J~\text{for some positive integer}~J$) can be found using
\begin{equation*} 
\mathbf{d}_{j} = \mathcal{D}_{0}\mathcal{G}(\mathcal{D}_{0}\mathcal{H})^{J-j-1}X,
\end{equation*}
for $j=0,\hdots,J-1$. Similarly, the scaling or smooth coefficients of the DWT are given by
\begin{equation*} 
\mathbf{c}_{j} = (\mathcal{D}_{0}\mathcal{H})^{J-j}X,
\end{equation*}
for $j=0,\hdots, J$. 

The information contained in the original time series $X$ can thus be fully described by the set of coefficients $ \{\mathbf{d}_{J-1}, \mathbf{d}_{J-2},\hdots,\mathbf{d}_{0},\mathbf{c}_{0}\} $. 

\paragraph{The Nondecimated Discrete Wavelet Transform (NDWT).} The {\em nondecimated} wavelet transform (NDWT) is a modification of the DWT outlined above in which the decimation step is not carried out, resulting in $2^{J}$ smooth and detail coefficients at each level of the transform. This allows for a fuller description of the local characteristics of the data in the decomposition, a feature that turns out to be particularly helpful for describing time series.  A more detailed treatment of the NDWT can be found in \cite{NSDWT}, see also \cite{coifman95:translation, percival95:on}. \bcb In the context of streaming data, the transform is such that only a small number of coefficients need to be recomputed at each time step, recycling previously evaluated coefficients.  This computationally efficient algorithm will be used within our online dynamic classification technique described in Section \ref{sec:method}. \ecb

\subsection{Multivariate locally stationary wavelet (MvLSW) processes} \label{subsec:MvLSWmodel}

We now turn to consider the application of wavelets within non-stationary time series models. Specifically we focus on the recently proposed multivariate locally stationary wavelet (MvLSW) framework introduced by \cite{Park}, which we later use to model the DAS data described in Section \ref{intro}. This approach provides a flexible model for multivariate time series that is able to capture (second order) nonstationarity, as well as temporally inhomogeneous dependence structure between channels of a multivariate series.  

Following \cite{Park}, a $P$-variate locally stationary wavelet process $\{\mathbf{X}_{t,T}\}_{t=1}^{T} $ can be represented as
\begin{equation} 
\mathbf{X}_{t,T}=\displaystyle\sum_{j=1}^{\infty} \displaystyle\sum_{k} \mathbf{V}_{j}(k/T)\psi_{j,t-k}\mathbf{z}_{j,k}, \label{MvLSWeq} 
\end{equation}
where $T=2^{J} \geq 1$ and $\mathbf{V}_{j}(k/T)$ is the lower-triangular transfer function matrix.  Each element of the transfer function matrix is assumed to be a Lipschitz continuous function with Lipschitz constants, $L_{j}$, that satisfy $\sum_{j=1}^{\infty} 2^{j}L_{j}^{(p,q)} < \infty$ for each pair of channels $(p,q)$.  The vectors $\psi_{j}=(\psi_{j,0},\hdots,\psi_{j,(N_{j}-1)})$ are discrete non-decimated wavelets associated to a low-\ / high-pass filter pair, $\{\mathcal{H}, \mathcal{G}\}$, constructed according to~\cite{Nason00} as
\begin{eqnarray*} 
\psi_{1,n}&=&\sum_{k} g_{n-2k} \delta_{0,k}=g_{n}\quad \text{for}~n=0,1,\hdots,N_{1}-1, \\
\psi_{j+1,n}&=&\sum_{k} h_{n-2k} \psi_{j,k} \qquad \quad \text{for}~n=0,1,\hdots,N_{j+1}-1.
\end{eqnarray*} 
In the equations above, $\delta_{0,k}$ is the Kronecker delta function and $ N_{j}=(2^{j}-1)(N_{h}-1)+1 $ where $N_{h}$ is the number of non-zero elements of the filter $\mathcal{H}=\{h_{k}\}_{k\in\Z}$. The random vectors $\mathbf{z}_{j,k}$ in \eqref{MvLSWeq} are defined such that $ E(\mathbf{z}_{j,k})=\mathbf{0}$ and $\text{cov}\big(z_{j,k}^{(i)},z_{j',k'}^{(i')}\big)=\delta_{i,i'}\delta_{j,j'}\delta_{k,k'}$. 

The local wavelet spectral (LWS) matrix and the wavelet coherence of a multivariate signal are key quantities of interest in the dynamic classification problem. Given a MvLSW signal $\mathbf{X}_{t}$ with associated transfer function matrix, $\mathbf{V}_{j}(z)$, the local wavelet spectral matrix $\mathbf{S}_{j}(z)$ is defined as
\begin{equation} \label{eq:spec}
\mathbf{S}_{j}(z)=\mathbf{V}_{j}(z)\mathbf{V}_{j}(z)^{\top}. 
\end{equation}
This quantity describes the cross-covariance between channels at each scale and (rescaled) location $z$.  The coherence is a measure of the dependence between the channels of a multivariate signal at a particular time and scale. Following~\cite{Park}, the wavelet coherence matrix $\boldsymbol{\rho}_{j}(z)$ is given by
\begin{equation}
\boldsymbol{\rho}_{j}(z)=\mathbf{D}_{j}(z)\mathbf{S}_{j}(z)\mathbf{D}_{j}(z), \label{wavecoh} 
\end{equation}
where $\mathbf{S}_{j}(z)$ is the LWS matrix from \eqref{eq:spec} and $\mathbf{D}_{j}(z)$ is a diagonal matrix with entries given by $S_{j}^{(p,p)}(z)^{(-1/2)}$.  It is these spectral and coherence quantities which we use to enable us to accurately classify multichannel signals with time-varying dependence and second order structure.

\paragraph{Estimation of MvLSW spectral and coherence components.}  In practice, the coherence and LWS matrix are unknown for an observed multivariate series and need to be estimated.  The LWS matrix of a multivariate signal can be estimated by first calculating the empirical wavelet coefficient vector 
$\mathbf{d}_{j,k}=\sum_{t} \mathbf{X}_{t}\psi_{j,k-t}$ at locations $k$ and scales $j$. 
The raw wavelet periodogram matrix is then defined as $\mathbf{I}_{j,k}=\mathbf{d}_{j,k}\mathbf{d}_{j,k}^{\top} $. 

\cite{Park} establish that the raw wavelet periodogram is a biased and inconsistent estimator of the true LWS matrix, $ \mathbf{S}_{j}(z)$. However, they show that (asymptotically) this bias is described by the well known inner product matrix of discrete autocorrelation wavelets, $\mathbf{A}$. The elements of $\mathbf{A}$ are given by $ A_{j\ell}= \sum_{\tau} \Psi_{j}(\tau)\Psi_{\ell}(\tau) $ where $ \Psi_{j}(\tau)=\sum_{k} \psi_{jk}(0) \psi_{jk}(\tau) $ (see \cite{Eckley} or \cite{Nason00} for further information). The bias inherent within the raw wavelet periodogram can therefore be removed using the inverse of this inner product matrix. To obtain consistency, the resulting estimate must be smoothed in some way, for example using a rectangular kernel smoother \citep{Park}. This results in an (asymptotically) unbiased, and consistent, estimator of the LWS matrix, $\mathbf{S}_{j,k}$, given by $ \hat{\mathbf{S}}_{j,k}= (2M+1)^{-1} \sum_{m=k-M}^{k+M} \sum_{l} A_{jl}^{-1}\mathbf{I}_{lm} $, where $M$ 
denotes the kernel bandwidth corresponding to a smoothing window of length $2M+1$.  Estimation of the wavelet coherence matrix is then straightforward, simply using a plug-in estimator, substituting $\hat{\mathbf{S}}$ into Equation~\eqref{wavecoh}.

With the key modelling notation established, we now briefly summarise an approach to dynamic classification based upon the MvLSW framework. This will be the cornerstone of the approach that we propose in Section 3. 

\subsection{Dynamic Classification}\label{subsec:DC}

Following their work on MvLSW processes, \cite{ParkDC} introduced an approach to dynamically classify a Multivariate Locally Stationary Wavelet signal $\mathbf{X}_{t}$ whose class membership may change over time.  The approach assumes that at any time $t$, the signal $\mathbf{X}_{t}$ can belong to one of $N_{c} \geq 2$ different classes, where $N_{c}$ is known. Let $C_{X}(t)$ denote the class membership of $X_{t}$ at time $t$ where $C_{X}(t) \in \{1,2,\hdots, N_{c}\}$. Following \cite{ParkDC}, the signal $\mathbf{X}_{t}$ in \eqref{MvLSWeq} can be then represented as
\begin{equation*} 
\mathbf{X}_{t}=\displaystyle\sum_{j} \displaystyle\sum_{k} \displaystyle\sum_{c=1}^{N_{c}} \mathbb{I}_{c}[C_{X}(k)]\mathbf{V}_{j}^{c}\psi_{j,k}(t)\mathbf{z}_{j,k}, 
\end{equation*}
where $\mathbf{V}_{j}^{c}$ is the class specific transfer function which is constrained to be constant over time and  $\mathbb{I}_{c}[C_{X}(k)]$ represents an indicator function which equals 1 if $C_{X}(k)=c$ and 0 otherwise.

To classify the multivariate signal $\mathbf{X}_{t}$, the approach makes use of a set of $N_{i}$ training signals, denoted $\big\{Y_{t}^{(i)}\big\}$ for $i \in \{1,2,\hdots,N_{i}\}$. It is assumed that the class membership of these training signals over time, $C_{Y^{(i)}}(t)$, is known. For each training signal, the LWS matrix $\hat{\mathbf{S}}_{jk;Y^{(i)}}$ and coherence matrix $\boldsymbol{\rho}_{jk;Y^{(i)}}$ can be estimated, as discussed in Section \ref{subsec:MvLSWmodel}.

The aim of this classification method is to calculate the probability of the signal belonging to a particular class at a given time point. To do this, the likelihood of the signal belonging to each class given the information contained in the training signals is calculated. It is necessary to apply a Fisher-z transform to the coherence estimates to ensure that the estimates can be approximated by a Gaussian distribution. For a class $c$, the transformed coherence $\zeta_{j}^{(c)}$ is given by
\begin{equation} \label{eq:fisher}
\zeta_{j}^{(c)}=\text{tanh}^{-1}\rho_{j}^{(c)}. 
\end{equation}
The mean and variance of the transformed coherence for class $c$ can be estimated using the transformed coherence for the training signals that are known to belong to that particular class. \bcb Note that in practice, the Gaussian distribution will be an approximation to the true distribution of the (finite sample) Fisher z-transformed coherence estimates. We recommend that, as for any such analysis,  this assumption is validated for any data set analysed. \ecb

As in \cite{Krz}, classification is performed using a subset of wavelet coefficients that show the largest difference between the classes in terms of the transformed coherence. The subset, denoted by  $\mathcal{M}$, consists of the scale and channel indices $(j,p,q)$ for $p<q$. $\mathcal{M}$ is made up of the coefficients that have the largest values of the discrepancy measure $\bigtriangleup_{j}^{(p,q)}$ given by
\begin{equation} 
\bigtriangleup_{j}^{(p,q)}=\displaystyle\sum\limits_{c=1}^{N_{c}} \displaystyle\sum\limits_{g=c+1}^{N_{c}} \Biggl\lvert \frac{\widehat{\zeta}_{j}^{(p,q)(c)}-\widehat{\zeta}_{j}^{(p,q)(g)}}{\sqrt{\text{var}\big(\widehat{\zeta}_{j}^{(p,q)(c)}\big)+\text{var}\big(\widehat{\zeta}_{j}^{(p,q)(g)}\big) }} \Biggr\rvert. \label{disind}
\end{equation}
In practice, a proportion $\wp$ is typically chosen and the subset $\mathcal{M}$ are those $\wp$\% of time-scale indices with the largest discrepancies \citep{Krz}.
In order to estimate the probability that the signal $\mathbf{X}_{t}$ belongs to a particular class at a given time, the transformed coherence $\widehat{\zeta}_{jk;X}$ is first estimated, before using Bayes' theorem to obtain
\begin{equation}\label{eq:bayes}
\text{Pr}\Big[C(k)=c \bigl\lvert\widehat{\zeta}_{jk;X}\Big] \propto \text{Pr}\big[C(k)=c\big]\mathcal{L}\Big(\widehat{\zeta}_{jk;X}\bigl\lvert\zeta_{j}(k/T)=\xi_{j}^{(c)} \quad \forall j \Big),
\end{equation}
where $\mathcal{L}(\theta |x)$ is the likelihood and $ \text{Pr}\big[C(k)=c\big]$ is a prior probability \citep{ParkDC}. Due to the use of the Fisher-z transform in \eqref{eq:fisher}, the likelihood $\mathcal{L}(\theta |x)$ takes the form of a Gaussian likelihood with mean vector $\mu^{(c)}$ and covariance matrix $\Sigma^{(c)}$. The vector $\mu^{(c)}$ consists of the elements of $\widehat{\zeta}_{j}^{(p,q)(c)}$ $\forall~(j,p,q) \in \mathcal{M}$, whilst similarly $\widehat{\mu}_{k}$ contains the elements~$\widehat{\zeta}_{jk;X}^{(p,q)}$~$\forall~j,p,q \in \mathcal{M}$. The Gaussian likelihood hence takes the form
\begin{equation} \mathcal{L} \Big(\hat{\zeta}_{jk;X}|\hat{\zeta_{j}}\big(k/T\big)=\zeta_{j}^{(c)}~\forall~j\Big)\propto \big|\Sigma^{(c)}\big|^{-\frac{1}{2}}\text{exp}\Big(-\tfrac{1}{2}\big\{\big(\hat{\mu}_{k}-\mu^{(c)}\big)^{\top}\big(\Sigma^{(c)}\big)^{-1}\big(\hat{\mu}_{k}-\mu^{(c)}\big)\big\}\Big).\end{equation}
In practice, the true mean vectors and covariance matrices of $\widehat{\zeta}_{jk;X}$ are unknown, and they are estimated using the training data. In the examples provided in Section \ref{sec:examples}, we use a flat (uninformative) prior.  However, of course, many other prior specifications could be used in the formulation above to reflect beliefs from application-specific expert knowledge.  

The dynamic classification method described here is an offline approach that calculates the probability of belonging to a particular class at each time point. Since we are interested in detecting stripes in DAS data in an online setting, we adapt the existing method to allow for classification of data streams. We describe our approach below.

\section{Online dynamic classification of multivariate series} \label{sec:method}

In order to adapt the existing dynamic classification method outlined in Section~\ref{subsec:DC} to an online setting, we make use of a moving window approach. The use of such a window encapsulates the constraint in many data streaming applications that there is only a limited data storage and memory with which to perform analysis.  

Our online dynamic classification technique proceeds as follows.  For a window of length $w=2^{J}$ the first step of our algorithm is to calculate the set of discriminative indices as defined in Equation~\eqref{disind} using a set of training signals of length $w$.  For reasons of efficiency, the discriminative indices are used in the classification step for each window of the data.  Although window-specific indices could be used, in our experience, updating the set of discriminative indices for each window increases computational complexity without providing significant accuracy improvement.  The dynamic classification method described in Section~\ref{subsec:DC} is applied to the first window of data to obtain the probability that the signal belongs to a particular class for the time points in the window.  

Upon arrival of a new data point, the window then shifts by one, and the data under analysis consists of the old data together with the new data point, but we also lose the first data point contained in the previous window. The online wavelet transform is then used to efficiently update the wavelet coefficients and the transformed coherence estimate for the new window. Using the information previously calculated from the training signals, we can then obtain the probability that the signal belongs to a particular class for the time points contained in the new window. The algorithm continues by repeatedly moving the window for each new data point and estimating the probability of each data point belonging to a class until we reach the end of the data stream. 

During our classification algorithm, we obtain multiple estimates for the probability that a signal belongs to a particular class (at each time point) from the different windows into which a data point falls. For example, for a time series of length $T$ analysed with a moving window of length $w<T$, we obtain $w$ estimates for the probabilities of an individual time point $t$ belonging to a given class $c$, which we denote $p_{t,i}^{(c)}$ for window $i$.

A question that arises as a result of the iterative approach is how to combine the estimates from different windows to obtain an overall probability that the time point belongs to a particular class, and hence classify the signal. In what follows, for computational simplicity we use a simple average, but other more sophisticated combination methods could be used.  In other words, our final probability estimates are given by
\begin{equation}
 p_{t}^{(c)}=\frac{1}{w}\sum_{i=1}^{w} p_{t,i}^{(c)} \qquad~\text{for}~t=1,2,\hdots,T. \label{average}
\end{equation}
In some applications, an overall classification of the signal is required rather than probability estimates. In this case, the class $c$ that has the largest probability $p_{t}^{(c)}$ is assigned to the time point $t$ for all $t \in \{1,2,\hdots,T\}$. 

A summary of our method for estimating the probability that a given multivariate signal belongs to a particular class $c$ at a particular time is given in Algorithm \ref{alg:onlineDC}.

\renewcommand{\figurename}{Alg.}
\def\pc{\value{figure}}
\setcounter{figure}{0}
\begin{figure}[!ht]
\begin{algorithm}{Online dynamic classification}
\begin{enumerate}
\item Let $X$ be a $P$-variate signal of length $T$ that we wish to classify using a moving window of length $w$.  
\item  Calculate the set of discriminative indices using a set of P-variate training signals of length $w$, whose class assignments are known.
 \item Apply dynamic classification method to the first window of data $X[~,1:w]$ to obtain the probability that the signal belongs to a particular class $c$ for the time points in the window, denoted $p_{t,1}^{(c)}$ for $t=1,2,\hdots,w$.
   \item Iterate for $i$ in $2$ to $T-w+1$
\begin{enumerate}[(a)]
 \item Apply the online wavelet transform to the new window of data $X[~,i:i+w-1]$ to update the wavelet coefficients.
 \item Update the transformed coherence at the set of discriminative indices using the wavelet coefficients calculated in the previous step.
 \item Apply dynamic classification method to obtain the probability that the signal belongs to a particular class for window $i$,  denoted $p_{t,i}^{(c)}$ for $t=i,i+1,\hdots,i+w-1$.
\end{enumerate}
\item Average probability estimates for each window using \eqref{average} to obtain the final probability that the signal $X$ belongs to a particular class $c$ over time, $p_{t}^{(c)}$ for $t=1,2,\hdots,T$.
\end{enumerate}
\end{algorithm}
\caption{Finding the average probability that a multivariate signal belongs to a particular class over time.\label{alg:onlineDC}}
\end{figure}
\renewcommand{\figurename}{Fig.}		
\setcounter{figure}{\pc}	
\addtocounter{figure}{1}

\section{Synthetic Data Examples} \label{sec:examples}

We now turn to assess the performance of our proposed online dynamic classification approach. To this end, a simulation study is designed to test the ability of this wavelet-based appproach to classify data streams exhibiting various characteristics.  More specifically, the study consists of three different scenarios. These scenarios are chosen to mimic signals arising in practice: 
\begin{description}
\item Scenario 1: Signal of length 1024, short time segments of length 100 between changes in class, nine class changes in total.
\item[]
\item Scenario 2: Signal of length 1024, alternating long/short segments of length 300 and 100 between changes, five class changes in total.
\item[]
\item Scenario 3: Signal of length 2048, long segments of length 300 between changes, six class changes in total.
\end{description}

For all scenarios, the generated series randomly switch classes between time segments.  A window length of 256 is used when implementing the online dynamic classification method and the training data consists of 10 signals, some of which contain changes in class. The R packages \textbf{wavethresh} \citep{wavethreshpackage} and \textbf{mvLSW} \citep{mvLSWpackage} are used to calculate the wavelet coefficients and transformed coherence that are used in the online dynamic classification.

Long segments of length 300 between class changes are chosen to ensure that there is a maximum of one class change in each dynamic classification window. In the situation where the class changes are reasonably far apart, we expect the online dynamic classification algorithm to classify the signal well. As a contrast, short segments of length 100 are also chosen to demonstrate some potential limitations of the method. In particular, when the signal contains multiple class changes that are close together, there is a possibility that our approach will misclassify the signals. \\

\begin{figure}[!htbp] 
\centering 
\begin{subfigure}[h]{0.8\textwidth}
 \includegraphics[width=104mm]{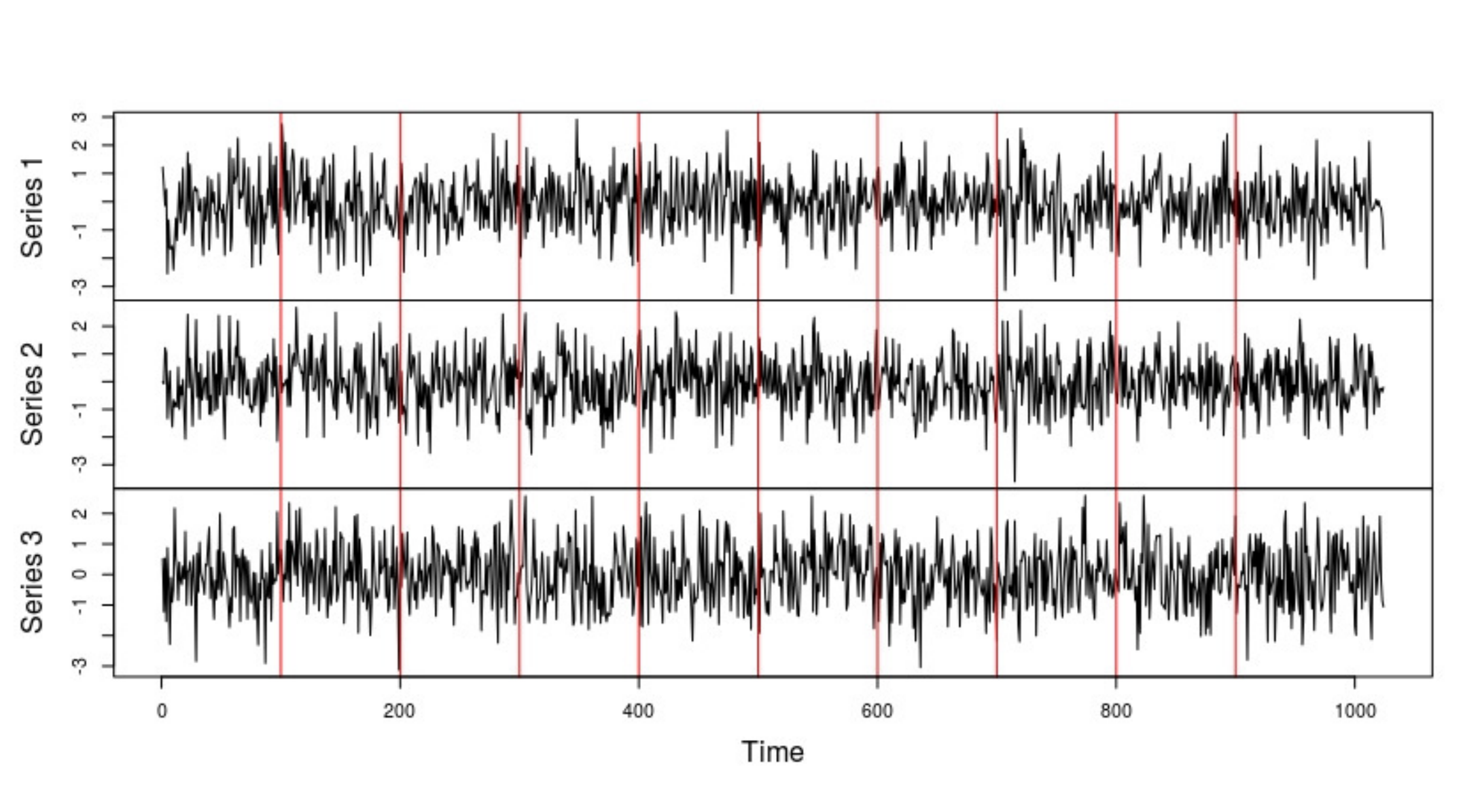} \caption{Scenario 1, multivariate Gaussian series} \label{fig:scen1} \end{subfigure} 
 \begin{subfigure}[h]{0.8\textwidth}
 \includegraphics[width=104mm]{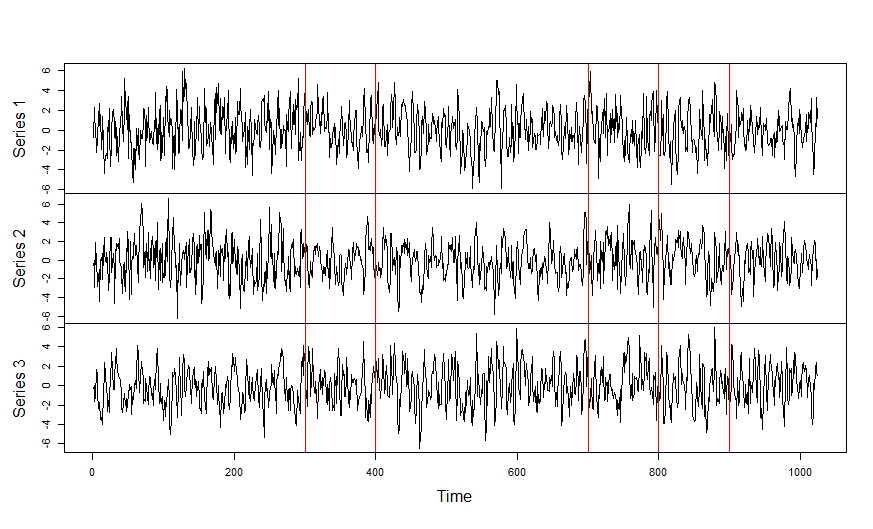} \caption{Scenario 2, vector moving average series} \label{fig:scen2} 
 \end{subfigure}
  \begin{subfigure}[h]{0.8\textwidth}
 \includegraphics[width=104mm]{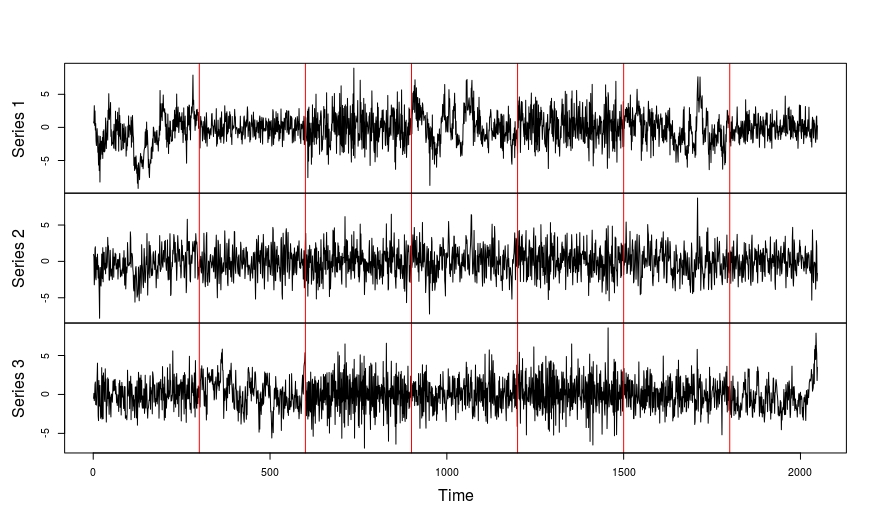} \caption{Scenario 3, vector autoregressive series} \label{fig:scen3} 
 \end{subfigure}
\caption{Example realisations of generating processes for the different scenarios used in the simulation study. (a) Short segments of length 100 between class changes; (b) Alternating long/short segments of length 300 and 100 between changes; (c) Long segments of length 300 between changes.}
\label{fig:scen}
\end{figure}

For each scenario, we consider a number of examples of generating processes for the classes in the multivariate series.   The first example we examine consists of three classes where each class is defined by a trivariate normal signal with mean $\mu=(0,0,0)$ and differing cross-channel dependence structure. More specifically, the classes are defined by the three covariance matrices 
$$\Sigma^{(1)}=
 \begin{pmatrix}
  1 & 0 & 0.3 \\
  0 & 1 & 0.7 \\
  0.3 & 0.7 & 1
 \end{pmatrix}\!, \Sigma^{(2)}=
 \begin{pmatrix}
  1 & 0.6 & 0.1 \\
  0.6 & 1 & -0.4 \\
  0.1 & -0.4 & 1
 \end{pmatrix} \, \textrm{and}\, \Sigma^{(3)}=
 \begin{pmatrix}
  1 & -0.5 & -0.2 \\
  -0.5 & 1 & 0.1 \\
  -0.2 & 0.1 & 1
 \end{pmatrix}.$$
Example simulated data for this process using the different class switching scenarios above are shown in Figure~\ref{fig:scen1}. 

To investigate the potential of our proposed approach further, we studied an example with a time-varying moving average (VMA) process, with three classes defined by the following coefficient matrices:\\

Class 1: \ $
\mathbf{X}_t = \mathbf{Z}_{t} + 
 \begin{pmatrix}
  1 & 0 & 0.6 \\
  0 & 1 & 0.3 \\
  0.6 & 0.3 & 1
 \end{pmatrix} \mathbf{Z}_{t-1} + 
 \begin{pmatrix}
  1 & 0.2 & 0.9 \\
  0.2 & 1 & 0.5 \\
  0.9 & 0.5 & 1
 \end{pmatrix} \mathbf{Z}_{t-2}
$

Class 2: \
$ \mathbf{X}_t = \mathbf{Z}_{t} + 
 \begin{pmatrix}
  1 & -0.7 & -0.3 \\
  -0.7 & 1 & 0.4 \\
  -0.3 & 0.4 & 1
 \end{pmatrix}\! \mathbf{Z}_{t-1} + 
 \begin{pmatrix}
  1 & 0.9 & -0.3 \\
  0.9 & 1 & 0 \\
  -0.3 & 0 & 1
 \end{pmatrix}\! \mathbf{Z}_{t-2},
$

Class 3: \
$ \mathbf{X}_t = \mathbf{Z}_{t} + 
 \begin{pmatrix}
  1 & -0.4 & 0.2 \\
  -0.4 & 1 & -0.6 \\
  0.2 & -0.6 & 1
 \end{pmatrix}\! \mathbf{Z}_{t-1} + 
 \begin{pmatrix}
  1 & 0.1 & -0.5 \\
  0.1 & 1 & -0.3 \\
  -0.5 & -0.3 & 1
 \end{pmatrix}\! \mathbf{Z}_{t-2},
$\\

where $\mathbf{Z}_{t}$, $\mathbf{Z}_{t-1}$ and $\mathbf{Z}_{t-2}$ are IID multivariate Gaussian white noise (see Figure~\ref{fig:scen2}).\\

The third example we consider is a vector autoregressive process with intra- and cross-channel changes in dependence between each class (see Figure~\ref{fig:scen3}).  The three classes in the example are defined by \\

Class 1: \ $
\mathbf{X}_t =
 \begin{pmatrix}
 0.2 & 0.3 & 0 \\
  0.3 & 0.5 & 0 \\
  0 & 0 & 0
 \end{pmatrix} \mathbf{X}_{t-1} + 
 \begin{pmatrix}
  0.6 & -0.1 & 0 \\
  -0.1 & -0.3 & 0 \\
  0 & 0 & 0
 \end{pmatrix} \mathbf{X}_{t-2} + \epsilon_{1},
$

Class 2: \
$ \mathbf{X}_t =
 \begin{pmatrix}
  0 & 0 & 0 \\
  0 & 0.4 & -0.4 \\
  0 & -0.4 & 0.4
 \end{pmatrix} \mathbf{X}_{t-1} + 
 \begin{pmatrix}
  0 & 0 & 0 \\
  0 & -0.6 & 0.2 \\
 0 & 0.2 & 0.3
 \end{pmatrix} \mathbf{X}_{t-2} + \epsilon_{2},
$

Class 3: \
$ \mathbf{X}_t =
 \begin{pmatrix}
  -0.1 & 0 & 0.4 \\
  0 & 0 & 0\\
  0.4 & 0 & -0.5
 \end{pmatrix} \mathbf{X}_{t-1} + 
 \begin{pmatrix}
  0.2 & 0 & -0.2 \\
  0 & 0 & 0 \\
 -0.2 & 0 & -0.3
 \end{pmatrix} \mathbf{X}_{t-2} + \epsilon_{3},
$\\\\
where the noise vectors $\epsilon_i$ are zero-mean multivariate normal realisations, distributed with covariances \\
$$ \Sigma_{\epsilon_1} = \begin{pmatrix}
  3 & 0.3 & 0.9 \\
  0.3 & 3 & 1.4 \\
  0.9 & 1.4 & 3
 \end{pmatrix}, \ \Sigma_{\epsilon_2} = \begin{pmatrix}
  2 & 1.3 & 0.4 \\
  1.3 & 1.8 & 0.3 \\
  0.4 & 0.3 & 2
 \end{pmatrix}, \ \Sigma_{\epsilon_3} = \begin{pmatrix}
  5 & 3.3 & 2.5 \\
  3.3 & 4.5 & 2.8 \\
 2.5 & 2.8 & 3.5
 \end{pmatrix}.$$

\paragraph{Competitor methods.}

In the simulation study, we compare our proposed method with a number of alternative classification techniques.  Firstly we consider a Hidden Markov Model (HMM) approach -- a probabilistic model of the joint distribution of observed variables, together with their ``hidden'' states (in this setting, classes).  Such methods have previously been used for classification in the literature, see for example \cite{HMMclass}.  In this model, it is assumed that (i) the observed data at a particular time is independent of all other variables, given its class and (ii) given the previous class, the class at a time is independent of all other variables (i.e.\ the changes in class are Markovian).  This means that we assume that the probability of changing class does not depend on time or previous class membership, which can be an unrealistic assumption to make in practice.  Furthermore, HMMs can be computationally intensive to implement especially in multiclass settings, requiring procedures such as the EM algorithm for 
tractable model fitting, see e.g. \cite{cappe09:inference}. An introduction to HMMs and their applications can be found in \cite{Zucchini}. 

A sequential HMM approach is applied to both the full test signal and its transformed coherence at the set of discriminative indices, using the {\em R} package \textbf{HMM} \citep{hmmpackage}. In both cases, \bcb the model is initialized to have equal state probabilities, and then trained using the initial data. When a new data point arrives, \ecb the probabilities of belonging to each state are computed. This process of increasing the number of data points and computing the probabilities is repeated until we reach the end of the signal. As with the online dynamic classification approach, multiple estimates for the probability of belonging to a state at a particular time point are obtained.  
This is because each time a data point falls within a window, probabilities associated to the time point belonging to a particular class are calculated. For each time point, the estimates are averaged and the overall classification of the signal is then defined to be the most likely state at each point.  We also considered a third variant of the sequential HMM approach that was applied to each window of the data used in the online dynamic classification and the corresponding transformed coherence. However this produced poor results so we omit them from the comparisons below. 

To demonstrate the importance of accounting for the dependence structure within the test series, we also apply a support vector machine (SVM) classifier to the series, available in the {\em R} package \textbf{e1071} \citep{e1071package}, as well as the mixture modelling approach from the \textbf{mclust} {\em R} package \citep{mclustpackage} (denoted {\em GMM}).   These methods do not explicitly allow temporal dependence in the classification rules, and so we would expect them to perform poorly in cases where this dependence features in the test series.   \bcb Specifically, we used a radial basis kernel for the SVM classifier. The GMM approach implemented allows for potentially different numbers of mixture components and covariance structures for each class, with the number of components chosen with the Bayesian information criterion (BIC). \ecb Similar to the HMM method described above, we show results on the SVM and GMM methods applied to the transformed coherence measure -- the results for the techniques 
on the raw series performed poorly and so they aren't reported in the tables.   In addition, we compare our method to the Na\"ive Bayes (NB) classifier in the \textbf{RMOA} \citep{RMOApackage} suite of online methods (again using the transformed coherence).  This latter technique uses a Bayesian 
classification rule similar to that in \eqref{eq:bayes}, and hence provides a useful comparison to our proposed use of time-varying wavelet coherence in a Bayesian rule. We also investigated the performance of several of the ensemble classification techniques implemented in the \textbf{RMOA} package, however their performance was similar to the NB classifier so we omit these results for brevity.

\paragraph{Training procedure details.} 
The training data for both the online dynamic classification and the sequential HMM approaches consists of ten signals of length 256. Of the ten signals, we simulate two each from Class 1, 2 and 3 and the remaining four signals contain a  mixture of all three. For the competitor methods that are applied to the transformed coherence measure, the training data has a slightly different form. In this case, the training signals are simulated with class memberships as defined above but the approaches are trained on the transformed coherences of these signals at the set of discriminative indices rather than the raw data. For the different scenarios and generating processes considered, in practice we find that the subset of most discriminative indices tends to consist of the finest scales, i.e.\ scales $1 - 3$, but that all channel indices appear to be important.

For each of the scenarios, 100 replications of the test signals are simulated and three different classification evaluation measures are considered. In particular, the number of class changes detected is recorded along with the V-measure \citep{Rosenberg} and the true positive detection rate, defined to be the proportion of each signal that is correctly classified. A change is detected if the signal switches class and this change lasts for longer than four time points. The V-measure assesses the quality of a certain segmentation (given the truth) and is measured on the $[0,1]$ scale where a value of $1$ represents perfect segmentation.  

The classification results for the different examples described above can be seen in Tables \ref{tab:mvn3} -- \ref{tab:ar1}.  Sequential HMM denotes the results for the full test signal and Embedded HMM denotes the results for the transformed coherence; similar descriptors are used for the SVM, GMM and NB classifiers applied to the transformed coherence of the raw data.  We remind the reader that these classification methods performed very poorly on the original series, and so are not reported in the tables.  In each case, we have recorded the average number of changes detected, V-measure and true positive rate (described above) over the 100 replications; the numbers within the brackets represent the standard deviation of the corresponding quantities. Recall that the number of true class changes for Scenarios 1, 2 and 3 are nine, five and six respectively.\\

\begin{table}[!h]
\caption{Performance of classification procedures over 100 replications of multivariate Gaussian series for different scenarios of class changes, using the evaluation measures described in the text.  Numbers in brackets represent the standard deviation of estimation errors. Bold numbers indicate  best result.}\label{tab:mvn3}
\begin{tabular}{cccc}
\hline\noalign{\smallskip}
 & Scenario 1 &  Scenario 2 &  Scenario 3 \\
 & (nine changes) & (five changes) & (six changes)\\
\noalign{\smallskip}\hline\noalign{\smallskip}
\multicolumn{1}{c}{Method} & \multicolumn{3}{c}{Average number of changes detected} \\
 \noalign{\smallskip}\hline\noalign{\smallskip}
Online dynamic classification ($w=256$) & \textbf{9.38 (0.65)} & \textbf{5.58 (0.88)} & 6.19 (0.44) \\
Sequential HMM  & 10.44 (3.80) & 6.71 (5.65) & 10.96 (13.78) \\ 
Embedded HMM  & 11.90 (3.29) & 8.49 (3.53) & 15.29 (3.74)  \\
Embedded SVM  & 13.55 (2.34) & 8.50 (2.02) & 8.02 (1.56) \\
Embedded GMM  & 17.88 (3.58)  & 15.45 (3.10) & 29.16 (5.34)  \\
Embedded NB  & 12.71 (2.27) & 7.43 (1.50) & \textbf{6.16 (0.42)} \\
\noalign{\smallskip}\hline\noalign{\smallskip}
\multicolumn{1}{c}{Method} & \multicolumn{3}{c}{Average V-measure} \\
  \noalign{\smallskip}\hline\noalign{\smallskip}
Online dynamic classification ($w=256$) & 0.89 (0.02) & 0.89 (0.03) & 0.94 (0.01) \\
Sequential HMM  & \textbf{0.94 (0.05)} & \textbf{0.93 (0.09)} & 0.94 (0.09) \\ 
Embedded HMM  & 0.78 (0.05) & 0.74 (0.10) & 0.80 (0.04) \\
Embedded SVM  & 0.80 (0.03) & 0.82 (0.04) & 0.91 (0.02) \\
Embedded GMM & 0.81 (0.02) & 0.73 (0.04) & 0.75 (0.03)   \\
Embedded NB  & 0.82 (0.06) & 0.86 (0.03) & \textbf{0.95 (0.01)} \\
\noalign{\smallskip}\hline\noalign{\smallskip}
 \multicolumn{1}{c}{Method} & \multicolumn{3}{c}{Average true positive rate} \\
 \noalign{\smallskip}\hline\noalign{\smallskip}
Online dynamic classification ($w=256$) & 0.91 (0.02) & 0.94 (0.02) & \textbf{0.97 (0.01)} \\
Sequential HMM   & \textbf{0.93 (0.11)} & \textbf{0.95 (0.11)} & 0.94 (0.13) \\ 
Embedded HMM & 0.59 (0.09) & 0.64 (0.13) & 0.75 (0.10)  \\
Embedded SVM  & 0.70 (0.05) & 0.85 (0.05) & 0.95 (0.02) \\
Embedded GMM  & 0.57 (0.06) & 0.64 (0.05) & 0.79 (0.05)  \\
Embedded NB  & 0.75 (0.06) & 0.88 (0.04) & \textbf{0.97 (0.01)}\\
\noalign{\smallskip}\hline
\end{tabular}
\end{table}

For the three class multivariate normal example (Table \ref{tab:mvn3}), it can be seen that all three methods overestimate the number of changes detected. The online dynamic classification performs the best in terms of the average number of changes detected, only marginally overestimating the number of changes, and is competitive with other methods in terms of V-measure and average true positive rate. Both the sequential HMM approach and the Na\"ive Bayes classification rule perform well in this setting according to the V-measure and the average true positive rate.  However, we note here that the improvement over our proposed method is minimal considering the variability in the estimates. 


\begin{table}[!h]
\caption{Performance of classification procedures over 100 replications of vector moving average series for different scenarios of class changes, using the evaluation measures described in the text.  Numbers in brackets represent the standard deviation of estimation errors. Bold numbers indicate  best result.}\label{tab:ma2}
\begin{tabular}{cccc}
\hline\noalign{\smallskip}
 & Scenario 1 &  Scenario 2 &  Scenario 3 \\
 & (nine changes) & (five changes) & (six changes)\\
\noalign{\smallskip}\hline\noalign{\smallskip}
\multicolumn{1}{c}{Method} & \multicolumn{3}{c}{Average number of changes detected} \\
 \noalign{\smallskip}\hline\noalign{\smallskip}
Online dynamic classification ($w=256$) & \textbf{9.82 (1.10)} & \textbf{5.78 (0.93)} & \textbf{6.59 (0.87)} \\ 
Sequential HMM & 35.75 (7.38) & 33.75 (8.82)  & 77.24 (16.54) \\ 
Embedded HMM & 10.28 (3.71) & 8.69 (2.74)  & 14.57 (4.89)  \\
Embedded SVM  & 11.90 (1.90) & 5.96 (1.10) & 9.18 (2.28) \\
Embedded GMM & 13.31 (2.89) & 14.21 (3.32) & 18.09 (4.56) \\
Embedded NB & 12.07 (1.63) & 5.87 (0.68) & 11.38 (2.36) \\
\noalign{\smallskip}\hline\noalign{\smallskip}
\multicolumn{1}{c}{Method} & \multicolumn{3}{c}{Average V-measure} \\
  \noalign{\smallskip}\hline\noalign{\smallskip}
Online dynamic classification ($w=256$) & \textbf{0.87 (0.02)} & \textbf{0.89 (0.03)}  & \textbf{0.94 (0.01)} \\
Sequential HMM & 0.76 (0.04) & 0.66 (0.05)  & 0.64 (0.06) \\ 
Embedded HMM & 0.75 (0.07) & 0.73 (0.08)  & 0.79 (0.05) \\
Embedded SVM & 0.85 (0.02) & 0.88 (0.04)  & 0.87 (0.03) \\
Embedded GMM & 0.81 (0.02) & 0.71 (0.03) & 0.81 (0.03) \\
Embedded NB & 0.84 (0.02) & 0.86 (0.03) & 0.87 (0.03) \\
\noalign{\smallskip}\hline\noalign{\smallskip}
 \multicolumn{1}{c}{Method} & \multicolumn{3}{c}{Average true positive rate} \\
 \noalign{\smallskip}\hline\noalign{\smallskip}
Online dynamic classification ($w=256$) & \textbf{0.89 (0.03)} & \textbf{0.93 (0.02)}  & \textbf{0.97 (0.01)}\\
Sequential HMM & 0.54 (0.15) & 0.57 (0.14)  & 0.50 (0.15) \\ 
Embedded HMM & 0.62 (0.08) & 0.67 (0.11) & 0.68 (0.05)  \\
Embedded SVM & 0.83 (0.03) & 0.91 (0.03) & 0.92 (0.03) \\
Embedded GMM  & 0.70 (0.05) & 0.62 (0.05) & 0.72 (0.03) \\
Embedded NB & 0.81 (0.03) & 0.92 (0.02) & 0.90 (0.03) \\
\noalign{\smallskip}\hline
\end{tabular}
\end{table}

As we introduce dependence into the series, the distinction between our proposed method and its competitors becomes more marked.  For the moving average process (Table \ref{tab:ma2}), the performance of the online dynamic classification method improves as we increase the length of the segments between class changes; on the other hand, the sequential HMM procedure (as with the other competitors applied to the original series) cannot cope with the dependence in the data, drastically overestimating the number of changes in the data.


\begin{table}[!h]
\caption{Performance of classification procedures over 100 replications of vector autoregressive series for different scenarios of class changes, using the evaluation measures described in the text.  Numbers in brackets represent the standard deviation of estimation errors. Bold numbers indicate  best result.}\label{tab:ar1}
\begin{tabular}{cccc}
\hline\noalign{\smallskip}
 & Scenario 1 &  Scenario 2 &  Scenario 3 \\
 & (nine changes) & (five changes) & (six changes)\\
\noalign{\smallskip}\hline\noalign{\smallskip}
\multicolumn{1}{c}{Method} & \multicolumn{3}{c}{Average number of changes detected} \\
  \noalign{\smallskip}\hline\noalign{\smallskip}
Online dynamic classification ($w=256$) & \textbf{9.79 (0.96)} & \textbf{5.36 (0.66)}  & \textbf{6.64 (0.87)} \\
Sequential HMM & 12.03 (5.49) & 9.30 (4.46)  & 17.27 (6.44) \\ 
Embedded HMM & 13.11 (2.70) & 10.44 (3.16) & 17.83 (6.60) \\
Embedded SVM &  12.80 (3.14) & 7.59 (2.03)  & 12.37 (2.78) \\
Embedded GMM & 16.82 (3.04) & 6.32 (2.75) & 19.64 (4.40) \\
Embedded NB & 14.64 (3.13) & 6.87 (1.40) & 11.91 (2.62) \\
\noalign{\smallskip}\hline\noalign{\smallskip}
\multicolumn{1}{c}{Method} & \multicolumn{3}{c}{Average V-measure} \\
  \noalign{\smallskip}\hline\noalign{\smallskip}
Online dynamic classification ($w=256$) &  \textbf{0.87 (0.02)} & \textbf{0.89 (0.03)}  & \textbf{0.92 (0.02)} \\
Sequential HMM & 0.81 (0.08) & 0.73 (0.09) & 0.75 (0.07) \\
Embedded HMM & 0.79 (0.03) & 0.71 (0.08)  & 0.74 (0.06) \\
Embedded SVM & 0.78 (0.03) & 0.83 (0.05)  & 0.84 (0.03) \\
Embedded GMM & 0.75 (0.02) & 0.61 (0.07) & 0.73 (0.04) \\
Embedded NB & 0.78 (0.06) & 0.84 (0.04) & 0.84 (0.03) \\
\noalign{\smallskip}\hline\noalign{\smallskip}
 \multicolumn{1}{c}{Method} & \multicolumn{3}{c}{Average true positive rate} \\
 \noalign{\smallskip}\hline\noalign{\smallskip}
Online dynamic classification ($w=256$) & \textbf{0.89 (0.02)} & \textbf{0.95 (0.02)}  & \textbf{0.96 (0.01)}  \\
Sequential HMM  & 0.70 (0.11) & 0.69 (0.09) & 0.65 (0.11) \\
Embedded HMM & 0.59 (0.09) & 0.60 (0.10)  & 0.59 (0.10)  \\
Embedded SVM & 0.65 (0.05) & 0.89 (0.04)  & 0.89 (0.04) \\
Embedded GMM & 0.51 (0.04) & 0.45 (0.04) & 0.51 (0.05) \\
Embedded NB & 0.64 (0.06) & 0.89 (0.03)  & 0.89 (0.03) \\
\noalign{\smallskip}\hline
\end{tabular}
\end{table}

The online dynamic classification algorithm outperforms the competitors consistently for the autoregressive series, as shown in Table \ref{tab:ar1}.   More specifically, it classifies the changes well in terms of the V-measure and true positive rate, i.e. a low misclassification rate. Provided that the set of training data accurately represents the range of classes present, we would expect the dynamic classification approach to be able to correctly detect both the location of the changes and the classes involved. In contrast, whilst the comparative methods detect the location of class changes well resulting in high V-measure, they can struggle to identify which class the signal belongs to after the class change has occurred, resulting in a lower true positive rate (a higher overall rate of misclassification). This can potentially be a challenge if accurate detection of anomalous areas is important. 

In addition, note that in nearly all cases across the examples and scenarios, there is less variability in the evaluation measures using our proposed online dynamic classification (indicated by lower standard deviations).  We also note here that the use of the coherence measure improves the performance of all competitor methods, justifying its efficacy as a classification feature in many settings.  Crucially, 
we also found that the online dynamic classification approach was faster than HMM-based methods for longer time series.

\section{Case Study} \label{sec:case}

In the previous section we considered the efficacy of our approach against tried and tested examples. We now turn to consider an application arising from our collaboration with researchers working in the oil and gas industry. 

The general philosophy is to apply our online dynamic classification method to acoustic sensing data provided by an industrial collaborator, with the aim of detecting striping within these signals. The training data consists of ten quadvariate signals of length 4096 obtained from a subsampled version of an acoustic sensing dataset. The class assignments for each of the training signals have been decided by an industrial expert. The test signal is obtained from the same dataset and is a quadvariate signal of length 8192, unseen in the training signals.  
The test series exhibits autocorrelation as well as dependence between series (see Figure \ref{fig:stripecoh}).  Due to the zero-mean assumption of the mvLSW model, in practice we detrend the series before analysis by taking first order differences of each component series (see Figure \ref{fig:stripes}(b)). 

We apply the online dynamic classification approach with a moving window of length 4096 to the test signal. Based on the results from Section \ref{sec:examples}, for comparison the sequential HMM method is applied to the full test signal with the first 400 data points used to train a two-state model. We also apply the sequential HMM approach to the transformed coherence of the test signal, again training a two state model using the first 400 data points. Two-state models have been applied to demonstrate our belief that the acoustic sensing data contains areas of stable behaviour and striping.  

\begin{figure}[!htbp] 
\centering 
\begin{subfigure}[h]{0.8\textwidth}
 \includegraphics[width=104mm]{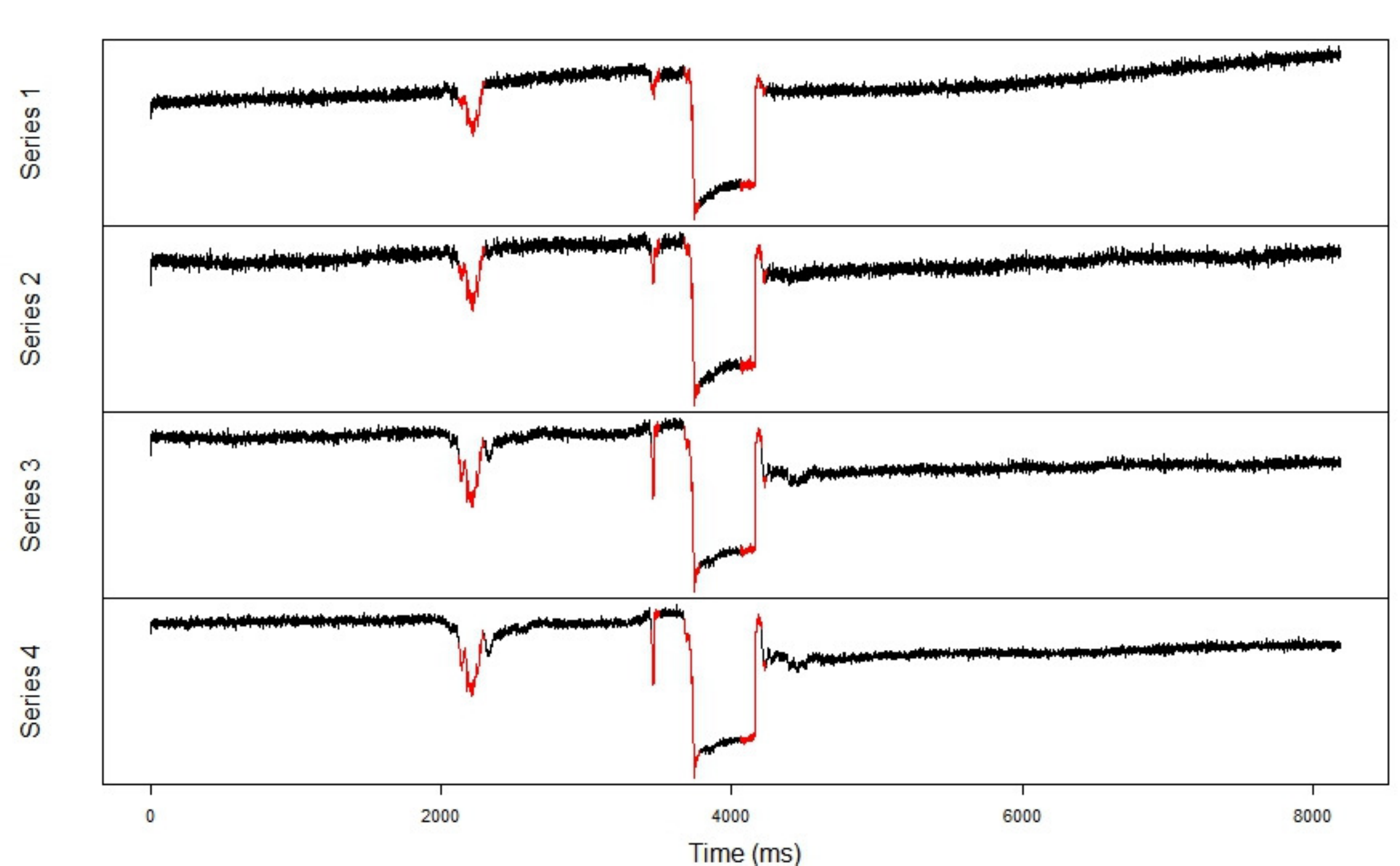} \caption{Online dynamic classification} \label{subfig-1:res1} \end{subfigure} 
 \begin{subfigure}[h]{0.8\textwidth}
 \includegraphics[width=104mm]{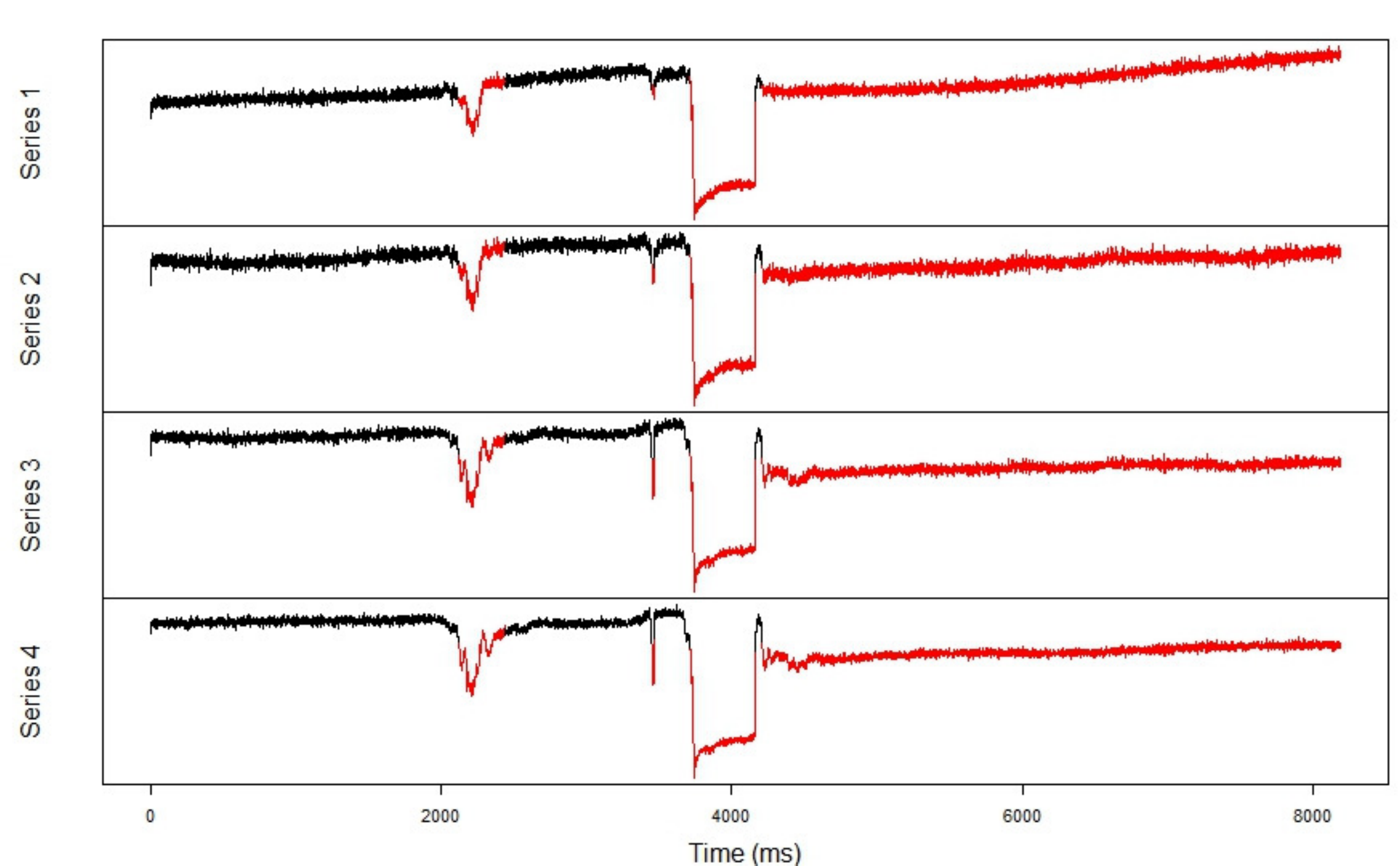} \caption{Sequential HMM} \label{subfig-2:res2} 
 \end{subfigure}
  \begin{subfigure}[h]{0.8\textwidth}
 \includegraphics[width=104mm]{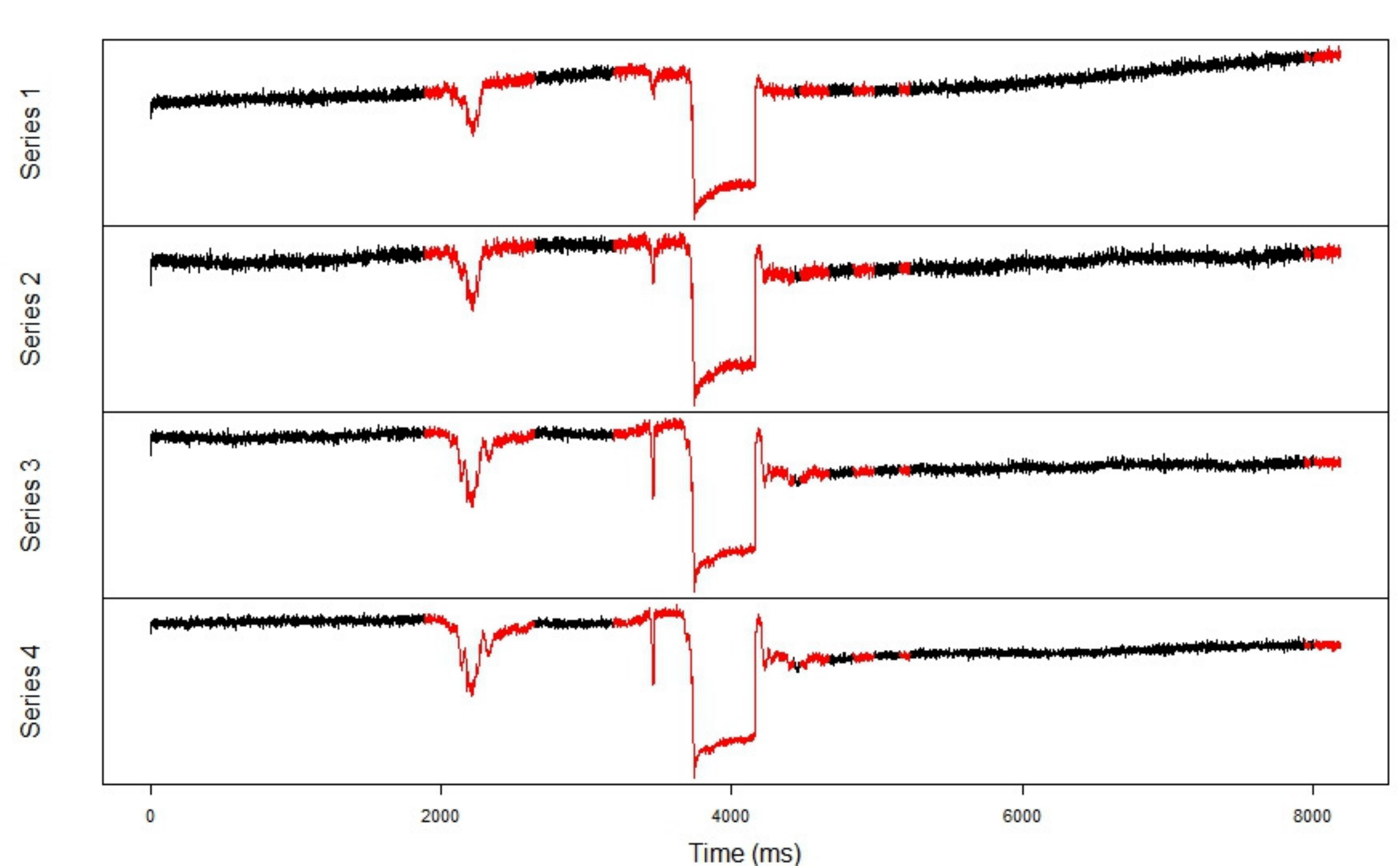} \caption{Embedded HMM} \label{subfig-3:res3} 
 \end{subfigure}
\caption{Classification results obtained from applying online dynamic classification, Sequential HMM and Embedded HMM approaches to acoustic sensing data, areas of the signal for which a change in class is detected are shown in red.}
\label{fig:res}
\end{figure}

In this case the true class membership of the test signal is unknown, therefore we compare the results visually. The classification results for each of the methods are found in Figure~\ref{fig:res}; areas of the test signal for which a change in class is detected are shown in red. It can be seen that the online dynamic classification method performs best in that it detects the stripes in the test signal with only minimal areas of misclassification when compared to the expert's judgement. In contrast, applying the sequential HMM approach to the transformed coherence of the signal also results in a change of class being detected at the stripes but the class changes take place over a longer period than we would expect. Finally, applying the sequential HMM method to the full signal results in the stripes being detected but the end of the test signal being misclassified. 

Recalling that the overall aim of this analysis has been to detect sudden regions of interest within (multivariate) acoustic sensing signals, as accurately as possible whilst minimising the number of falsely detected points -- then the results look very positive. Specifically the classification results obtained by the online dynamic classification method compare very favourably.   It is interesting to note that in these examples, coarser scales (i.e.\ scales 6-11) appear to play a key role in the classification. 

When compared with a subjective analysis of the data as displayed in Figure \ref{fig:stripes} we see that, each method correctly assigns `non-stripe' regions with the following (correct) classification proportions; 0.944 for online dynamic classification,  0.792 for embedded HMM and 0.477 for sequential HMM.
  
\section{Concluding remarks} \label{sec:conclusions}
In this article, we introduced an online dynamic classification method that can be used to detect changes in class within a data stream. We demonstrated the efficacy of the method using simulated data examples and an acoustic sensing dataset from an oil producing facility. The case study shows that our approach can be successfully used to detect anomalous periods in acoustic data, resulting in fewer areas of misclassification compared to more traditional classification methods, such as Markov Model approaches. Moreover, we have found that the use of a coherence measure in classification improves the performance of these methods.   

In practice, we have found that a parsimonious choice of window is required: as with other moving window approaches, too short a window, and the results are not satisfactory; too long a window increases the computational time and potentially produces edge effects. We leave the challenge of automatically choosing window length as an avenue for future research. In addition, we have observed that, as with other competitor methods, our approach classifies well when the distance between class changes is comparable to the window length but can struggle when we have shorter segments between changes.  

Future work may consider the problem of detecting stripes that are characterised by more gradual changes in their properties. In practice, these features may be less obvious and we might wish to not just detect but also classify the type of stripe present in the acoustic sensing data. Our method could potentially be used to do this, provided that our training data represents the range of stripes that we wish to classify.

\appendix
\section{Comparison of computational cost of online classification methods}\label{sec:computcost}

In this section, we provide an analysis of the computational cost of the various competitor classification methods outlined in Section \ref{sec:examples}.  To this end, we run each online classification method on a set of test signals of increasing length, namely $T=1024,2048,4096, 8192$.  In particular, for each method and series length, we record the runtime of each method, averaged over $K=25$ replications of series from the first example in Section \ref{sec:examples}.  This allows us to compare how the runtime of each method scales with series length, $T$, removing external factors such as efficiency of coding and implementation programming language. 

\begin{figure}[!htbp] 
\centering 
 \includegraphics[width=104mm]{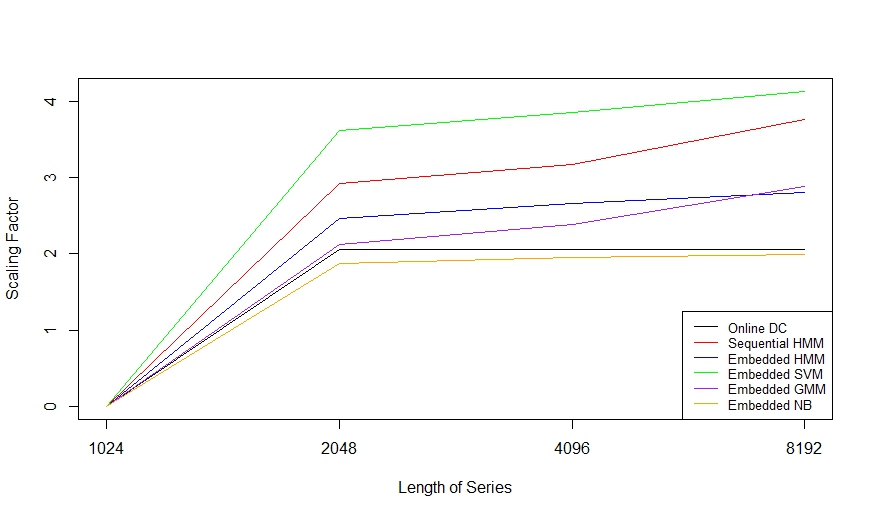}
\caption{Comparison of computational cost of the classification methods described in Section \ref{sec:examples} in terms of their scaling behaviour with length of test series.}
\label{fig:runtime}
\end{figure}

The results of the runtime analysis are shown in Figure~\ref{fig:runtime}.  As seen from the plot, as expected, each method increases in runtime with the length of the series.  However, after an initial increase, our dynamic online classification method has a desirable near constant scaling with the length of the series.  Its scaling profile is the best after the NB classifier.  Given the improvement in classification over the competitor methods across the range of examples studied in Section \ref{sec:examples}, we feel that this profile justifies the use of the proposed method.

\bibliographystyle{spr-chicago}      
\bibliography{draft}   
\nocite{*}

\end{document}